\documentclass[]{elsarticle}
\usepackage{times}
\usepackage{palatino}
\usepackage{amsmath}
\usepackage{amsfonts}
\usepackage{amssymb}
\usepackage{graphicx}
\usepackage{graphicx,color}
\usepackage{alltt}
\usepackage{mathrsfs}
\usepackage{epsfig}
\usepackage{subfigure}
\usepackage{verbatim}
\def\ll{\label}
\def\re{\ref}
\def\c{\cite}

\def\r1{(\ref{$1})}

\def\ba{\begin{array}{c}}

\def\ea{\end{array}}

\def\De{\Delta}

\def\l{\left}
\def\l({\left(}
\def\r){\right)}
\def\r{\right}

\def\la{\lambda}
\def\al{\alpha}

 \def\be{\begin{equation}}
\def\bc{\begin{center}}
\def\ec{\end{center}}
\def\bit{\begin{itemize}}
\def\eit{\end{itemize}}
\def\ee{\end{equation}}
\def\ed{\end{document}}
\def\bea{\begin{eqnarray}}
\def\eea{\end{eqnarray}}
\def\efr{\end{flushright}}

\begin{document}

\title{{ 
Exact Bethe ansatz solution for a 
  quantum  field model  of interacting scalar fields   
in quasi-two  dimensions
  }}
\author[]{Anjan Kundu} \ead{anjan.kundu@saha.ac.in}
 \address{ Saha Institute of Nuclear Physics, Theory Division,
 Kolkata, INDIA}
\vskip 1cm 




\date{\today}
\begin{abstract}
  Integrable quantum field models are known to exist mostly in  one
space-dimension.  Exploiting the concept of multi-time in integrable systems
and  a Lax matrix of
higher scaling order, we construct a novel  quantum field model in quasi-two
dimensions involving interacting fields. 
  The  Yang-Baxter integrability is proved for the model by  finding a new
kind of
commutation rule for
  its  basic fields,  representing  nonstandard  scalar fields along the
transverse direction.  
 In spite of a close link  with the  quantum
Landau-Lifshitz equation, the present  model differs widely from
it,
 in  its content and the  result obtained.
Using
further  the algebraic Bethe ansatz we solve
exactly the eigenvalue problem  of this quantum field model   
 for all its  higher  conserved operators.  The idea presented here
should instigate the construction of a novel class of integrable field and
lattice models and exploration of a new type of underlying algebras.
 \end{abstract}

\maketitle
\noindent {PACS numbers}:
{ 02.30.lk,
11.10.Kk,
03.65.Fd,
03.70.+k,
11.15.Tk
\\
 
\maketitle

\section{Introduction and Motivation}

 Quantization of  integrable    field models, in spite of their highly  nonlinear
interactions, and the exact  nonperturbative
solution of their eigenvalue problem through an algebraic generalization  of the
Bethe ansatz \c{bethe31}  (ABA) 
 was a real breakthrough in the theory
of quantum integrable systems \c{fadrev,kulskly,korbook,baxter}. 
Universal appeal of this approach was
understood, when the same construction was applied successfullly to a
number   of   quantum field  models of  diverse  nature, e.g.   
 nonlinear Schr\"odinger   (NLS) equation
\c{fadrev,sklyaninNLS} and derivative NLS equation \c{kunDNLS} field models,
  sine-Gordon \c{fadSG} and  Liuoville \c{fadLiu}  models, quantum
Landau-Lifshitz equation (LLE) \c{qLLE}
etc., apart from a rich family of quantum  lattice models 
\c{bethe31,xxz,toda,kunToda,tj,sklyGodin}. 
 It is also revealed, that this family of quantum integrable  models can be
 generated  from a single ancestor Lax matrix or its q-deformation,
 exploring the deep reason behind the boarder applicability of the
method \c{korbook,kunduPRL99}. 

Nevertheless, 
behind the success of this unifying scheme, there seems to be   a
limiting factor restricting the existing quantum integrable models 
within the structures defined by the   ancestor model Lax operator and 
confining  their construction only to  one space-dimension (1d).
The Kitaev models \c{Kitaeev12}, solvable   in two space-dimentions,
 though belong to a different class, seem to be 
rather exeptions. 

Recall, that the well known $(1+1) $-dimensional NLS equation 
\be iq_{t}=q_{xx}+2(q^*q)q, \ \ll{nls} \ee
 with  subscripts 
denoting partial derivatives,
extended as an integrable  quantum field  model
 involving bosonic   scalar field: 
 \be 
[ q(x,t), \ {q}^\dagger_x(x^{'},t) ]=-i  \  \delta(x-x{'}), \ 
  \  \ll{CRnls}\ee
was solved way back in eighties \c{fadrev,sklyaninNLS}. 
A recent proposal on the other hand 
constructs, going beyond the known ancestor model Lax operator,
 a new type of integrable  2d quantum NLS field model, using  
a higher order Lax matrice \c{KundNPB15}.
At the classical level this quasi-$(2+1) $ dimensional  NLS equation may be given by

\bea
iq_{t}&+&q_{xy}+2i(q^*_{x}q-q^*q_{x})q=0, 
\ll{NLS2} \eea
which differs significantly from the standard NLS (\ref{nls}).
As shown in   \c{KundNPB15},
 at the
quantum level, this quasi-2d NLS model represents a quantum integrable
system, where the basic complex scalar field of the model 
$ 
q(x,y,t), $     
satisfies an unusual commutation rule (CR): 
\be 
[ q(x,y,t), \ {q}^\dagger_x(x,y^{'},t) ]=-2i  \  \delta(y-y{'}), \ 
[ q(x,y,t), \ {q}^\dagger(x,y{'},t) ]= 0, 
  \  \ll{CRx}\ee
along the transverse direction $y$,  widely different from the bosonic CR (\ref{CRnls}).
 However, for establishing the  universality of this nonstandard
approach,
one needs at least another example of a  quantum integrable model,
where the scheme for  constructing integrable quasi-2d  field
models   could be applied. Untill the date no such proposal for new models,
integrable in  higher dimensions, exploiting the idea of Lax operator
of higher scaling orders \c{KundNPB15}  has been offered yet.

Our motivation here therefore, is  to construct  a significantly new type of
quantum  field model in qusi-two dimensional space, exploiring  the  
concept pf multi-time dimension in integrble systems and   following
the idea
of using a higher order Lax  operator  and at the same time 
 to solve the model exactly by the algebraic
Bethe ansatz. The proposed integrable model shows an intimate connection  
with the quantum LLE \c{qLLE}, though there are also wide differences.
 It is well known that quantum LLE is receiving renewed
interest in  recent years in connection with the string theory, since the
string states are found to be equivalent to its dual gauge theory,
represented by the effective LLE model, starting from  semiclassical to the
exactly solvable quntum level. Therefore, the proposed  field model with
a close link  to the  quantum LLE model might also be important from the
string theory point of view.
Moreover,  the underlying algebraic structure of the basic 
fields involved in our model, guaranting  
the  quantum integrability of the system, represents  a new fundamental quantum commutator
different from  all such  algebraic relations known for the  existing models. 

\section{Construction of the integrable field model:  classical case }

Recall that the LLE
 \be
{\bf S }_t=[{\bf S} , {\bf S }_{xx}], \ \  \ {\bf S }^2=I,  \ll{lle}\ee 
involving spin field ${\bf S} (x,t)=(S^1,S^2,S^3) $    with the known CR:
 \be
[{ S^a (x,t)}, S^b (x^{'},t) ]=\epsilon^{abc}  S^c(x,t) 
 \delta(x-x{'}),  \ll{CRlle}\ee 
is a $(1+1) $ dimensional    integrable system,
 both at the classical and the quantum level.
Classical LLE is gauge equivalent to the NLS equation \c{NLSgeLLE} and
similar to the NLS model the quantum LLE satisfies the Yang-Baxter equation
with rational $R$-matrix and is exactly solvable by the Bethe ansatz \c{qLLE}.
Integrable systems share the excluisive property of association with a Lax
operator, which with its several far reaching consequences, may be
considered as a strong criterion for the integrability of the model itself.
 The  space-Lax
operator  associated with the LLE:
\be U_{lle}(\la)=\frac 1 \la \  {\bf S}, \ \ {\bf S}^2=I \ll{Ulle} \ee
 represents an infinitesimal space-shift operator 
in the $x$-direction, ass0ciated   with the linear Lax equation   $\Phi_x=U_{lle}(\la)\Phi $
 and falls in
the standard structure of the rational ancestor model \c{kunduPRL99} with linear dependence on
the spectral parameter $ \frac 1 \la$ and on  the basic fields.    
However, for the present model
  we  look for a   Lax operator structure with
nonlinear dependence  on the spectral parameter as well as on the basic
fields. Such Lax operators, though   known in the
literature, mostly have  never  been used as a quantum Lax operator 
involved  in the quantum integrability and for  the construction of 
quantum model Hamiltonians.

\subsection{Lax operator}

Using the concept of multi-time dimension and the  space-time duality in integrable systems 
 investigated recently \c{KundNPB15,suris,duaNPBl16}, 
we use the time-Lax operator of the LLE system
\be 
U(\la )=\frac {2i} {\lambda^2} {\bf S} +\frac {i} {2 \lambda} ({\bf S} {\bf S}_x- {\bf S}_x  {\bf
S}) 
\ll{V}\ee
 and 
 define it  as a space-Lax operator along an additional  space dimention, 
 defining  $ U(\la )$ as a generator for the  
shift  along the $y$ direction: $\Phi_y=U(\la;x,y,t)\Phi ,  $  in a quasi-$(2+1) $ dimensional  integrable system.
Notice, that the  space  Lax operator (\ref{V})  is of higher scaling order compared 
to  $U_{lle}(\la )$   (\ref{Ulle}) for the LLE and other  known type
models. Since
  $x$ and $ \frac {1} {\lambda}$ go as the
length $ L$,
 the scaling dimension (SD)  of $U_{lle}(\la )$ and other AKNS type
models \cite{LLE} become  $1$, while for the Lax operator 
(\ref{V}) the SD  consequently result to $2$,  with  
$y$  scaling as $L^2$ . 
Note also, that in spite of 
 an intimate 
connection with the  the   $(1+1)
$-dimensional  standard  LLE model (\ref{lle}), the present  model, as we  see fbelow, differs
widely   from it, both at the classical and the quantum level.
The  current field components  
${\bf S}=(S^1,S^2,S^3) $ apeearing in (\ref{V}), unlike the 
spin field CR (\ref{CRlle}) exhibit  unusual  characteristics, expressed through
 novel   algebraic relations
 (\ref{PB}) , as a consequence of
the integrability of the system.  The present model is defied
 in a 
$(2+1)$- dimensional  space-time
$(x,y,t)  $ and   the related field CR (\ref{PB}), defined at different space points along
 the $y$ direction and   unlike (\ref{CRlle})
for the
LLE moedel, does not allow the usual  consraint $\ {\bf S }^2=I, $.   

 As predicted in \c{KundNPB15}, the  models  
with such Lax matrices of higher scaling order could constitute a new
family of  integrable systems satisfying the Yang-Baxter equation
with fundamentally new algebraic relations. We would find, that  the proposed     
 Lax matrix (\ref{V}) indeed constructs a ultralocal integrable model of
the rational class and though bears  resemblance with the LLE \c{qLLE},
it is defined in 
higher than 1d space dimensions and exhibits a new commutation relation,
significantly
different from the well known algebraic relations, like spin algebra, bosonic
algebra etc.. In
particular,
the  field components $ S^a, \ a=1,2,3 $ in the present quasi-$(2+1) $
dimensional nonlinear 
model 
  may be defined in the classical case, through the  Poisson bracket (PB)
relations as  
\be 
 \{ S^a(x,y,t),S^b_x(x,y',t) \}=   \frac {i} {2 }\delta_{ab} \delta(y -y')  
, \ \ a,b \in [1,2,3]\ll{PB}\ee
Looking closely to  relations (\ref{PB}) we can observe several
novelties.
First, these PB relations for the field $ {\bf S}$  are  defined at 
space points along the
$y$ direction and does not allow $ {\bf S}^2 $ as a Casimir operator, 
while those  in case of   the known  LLE model  are valid along the
$x$ direction and allows the constraint $ {\bf S}^2 =1 $. 
Second,
 the PB relations     involve $x$ derivative of the
fields, which goes beyond the known algebraic relations like for the  spin, boson,
fermion  etc. (Indeed we are familiar  with canonical brackets involving field and
its time-derivative only). Third, unlike the spin algebra (\ref{CRlle}),  
 nontrivial relations 
in  (\ref{PB}) exist
only between the same field components, which is same for all individual
field  components. Importantly, in spite
of the appearance of the derivative term in the Lax operator  (\ref{V})  
the specific form of  (\ref{PB}), as we see below, guarantees the
integrability of the system as an ultralocal model. Rewriting 
the PB relations   (\ref{PB}) as
\bea 
 \{ S^\pm(y,x,t),S^\mp_x(y',x,t) \}&=&   i \delta_{ab} \delta(y -y'), \
S^\pm=
S^1 \pm i 
S^2 \ll{s+-}\\
 \{ S^3(y,x,t),S^3_x(y',x,t) \}&=&   i  \delta(y -y'),
.\ll{S3}\eea
and comparing them with the relation (\ref{CRx}) for the nonstandard  complex scalar field 
 proposed recently in an integrable quasi-2d NLS model
\c{KundNPB15}, we notice
that  $S^+ $  and its conjugate  $S^+ $ satisfy a   relation similar 
to  the  nonstandard  complex  field, with  $S^3$ satisfying  also a similar     relation,
though  as a real scalar field.
It is interesting to note that, 
while the basic   fields  behave  like  spin fields  in  the   LLE model
with respect to the PB (\ref{CRlle}) stretching  along the $x$-direction and 
  are  expressible  through   bosonic scalar fields  
 through Holstein-Primakov
transformation,  in  the present model the basic fields  following the PB
relations (\ref{PB}) along the $y$-direction,   behave
like scalar fields  themselves,  
revealing   their identity  as a  complex 
 scalar  and a real scalar  field, with unusual PB relations as
(\ref{s+-},\ref{S3}). Another major difference between the present and the
known  
LLE model is, that   $  {\bf S}^2={S^1}^2+{S^2}^2+
+{S^3}^2\equiv s^2 $ is a Casimir operator for all the  components of the spin field 
in the PB (\ref{CRlle}) related to the LLE model and therefore  one can set
the constaint $s^2=1 , $ reducing the degrees of freedom of the fields to $2$,
 linking them to  a single  bosonic field. However, 
 for the present model with PB relation (\ref{s+-},\ref{S3})  the function $
{\vec S}^2=s^2(x,y,t)
$  is  no longer  a Casimir operator resulting no constraints and remains as a derived 
field, which consequently leaves the independent  degrees of 
 the current field to be  $3$, corresponding to a cpmplex scalar field
$S^+,S^-
$ together
with a real scalar field $S^3 $.  

\subsection{Hamiltonian and higher conserved quantities}
The Lax oprators $U(\lambda) $ in general  may be considered as  infinitesimal shift
generators along
different space-time  directions,  defining the associated linear system,
which for  (\ref{V}) takes the form   
\be 
\Phi_y(x,y,t, \lambda) =U(\la )\Phi (x,y,t, \lambda), \ \Phi = \left( \begin{array}{c}
 \phi_1  \\ \phi_2 
          \end{array}   \right), \ 
 \ll{Uy} \ee
where $U(\la )$  represents a shift operator along the $y$
direction, which is the relevant direction here, showing a quasi-2d nature
of our model. 
This system with explicit information about the Lax operator is rich enough
to  generate all higher conserved  quantities
including the  Hamiltonian of the model  and to  solve the related  
  hierarchy of  nonlinear equations through inverse scattering
technique (IST). Since we are concerned  about  the
quantum generalization of the model, we  do not deal here  with the
classical solutions of the  nonlinear equations through the IST
and are   interested   only in 
 the explicit construction of the conserved quantities     $  C_n, \
n=1,2,\cdots $, which may be derived from the Lax equation (\ref{Uy})  as 
\be 
  \ \ln \phi_1= i \sum_n C_n \lambda^n, \
C_n=\int dy \ \ \rho_n, 
\ll{phiCn}\ee
where for constructing the densities $ \rho_n,\ n=1,2,\cdots  $ of 
conserved quantities we may use the matrix elements $ U_{ij}$  of the Lax operator:  
\be 
i \sum_n \rho _n \lambda^n = U_{11}+U_{12} \ \Gamma, \ \ \  \Gamma= \frac  {\phi_1 }
{\phi_2} 
\ll{phiGam}\ee
to derive a Riccati equation of the orm
\be 
\Gamma_y = U_{21}-2U_{11} \ \Gamma- U_{12} \Gamma^2 , \Gamma= \sum_{n=0}
\Gamma_n  \lambda^n,
\ll{Riccati}\ee
where using the  expressions of  Lax matrix (\ref{V}) one gets
the 
recurrence equation

\be 
\Gamma_{ny}
= 4S^3 \ \Gamma_{n+2} + 2(S^-S^+_x-S^+S^-_x) \Gamma_{n+1}   
+ 2S^-\sum_{k=0}^{n+2} \Gamma_{n+2-k} \Gamma_{k} + (S^3S^-_x-S^-S^3_x) 
\sum_{l=0}^{n+1} \Gamma_{n+1-l} \Gamma_{l}
\ll{Riccatin}\ee
for $n>0 $ where $\Gamma_0=-\frac {S^3+{s}} {S^-},$ \ with $ \ s= \sqrt{{\bf
S}^2} $. Recall again that unlike the LLE model, here $ s(x,y,t)$  is not a constant
but a real field,
which makes all the field components $ S^1,S^2,S^3$ to be independent of
each other.  
Solving re4currence relations  (\ref{Riccatin}) one gets in the first step 
\be 
\Gamma_{1}= \frac 1 {4s} [(S^3S^-_x-S^-S^3_x)\Gamma^2_{0} +
(S^-S^+_x-S^+S^-_x) \Gamma_{0}   
 + (S^3S^+_x-S^+S^3_x)], \ \ \Gamma_0=-\frac {S^3+{s}} {S^-},
\ll{Gam1}\ee
\bea 
\Gamma_{2}= \frac 1 {2(2S^3+s)} [\frac 1 2 \Gamma_{0y}-(S^-S^+_x-S^+S^-_x)
\Gamma_{1} +S^-
\Gamma^2_{1}+ 
(S^3S^-_x-S^-S^3_x)\Gamma_{1} \Gamma_{0}   
,
\ll{Gam2}\eea
 etc.
Inserting these relations in (\ref{phiCn}-\ref{phiGam}) we can derive finally the                                                              
infinite set of commuting conserved quantities as 
\be 
C_{n-2}=\int dy ( 2S^-\Gamma_{n}+ (S^3S^-_x-S^-S^3_x)\Gamma_{n-1}, \ n\geq 2
.
\ll{Cn}\ee
Therefore  the lower order conserved quantities may be given in   the
 explicit form
\bea 
C_{-2}&=&\int dy \ 2 S^3+2S^-\Gamma_{0}=\int dy (S^3+s) , \nonumber \\  
C_{-1}&=&\int dy (  (S^-S^+_x-S^+S^-_x) + (S^3S^-_x-S^-S^3_x)\Gamma_{0}+
2S^-\Gamma_{1} \nonumber \\
C_{0}&=& \int dy ( (S^3S^-_x-S^-S^3_x)\Gamma_{1}+
2S^-\Gamma_{2}\nonumber \\
C_{1}&=&\int dy (  (S^3S^-_x-S^-S^3_x)\Gamma_{2}+
2S^-\Gamma_{3}   
\ll{Cm21012}\eea
etc. 
The  
Hamiltonian of the model can be  defined as 

\be 
H=C_{2}=\int dy (  (S^3S^-_x-S^-S^3_x)\Gamma_{3}+
2S^-\Gamma_{4}) 
\ll{H}\ee
where solutions for $ \Gamma_2,\Gamma_3, \Gamma_4$ calculated from 
recurrence relations (\ref{Riccatin}) using  (\ref{Gam1}), (\ref{Gam2})  
are to be inserted,  that are
straightforward but a bit lengthy, which  we omit  here.
  Notice the    quasi $(2+1) $-dimensional nature of the   Hamiltonian,
since though (\ref{H}) with (\ref{Riccatin}, \ref{Gam1}, \ref{Gam2})  involve
 both $x$ and $y$ derivatives of the
field,
the volume  integral is taken  only along the $y$ direction.
The space-asymmetry with   the   appearance of  space derivatives $ S^a_{x}(x,y,t)$ and
$S^a_{y}(x,y,t)$ in an  assymetric way 
is also explicit. 

\subsection{Classical Yang-Baxter equation}

For proving the complete integrability of a system it is not enough to have
all higher conserved quantities $C_n, \ n=1,2,\ldots $ , but one has to show that they are  all independent
entries i.e.,   are in involutions. Therefore, one has to  show that the conserved
quantities Poisson-commute $\{ C_n, C_m\}=0 $
 (operator commute for quantum models). For proving this global  statement for our model,
one may demand   a local sufficient  relation on the Lax matrix as   
\bea 
\{U(\la, x, y ), \otimes U(\mu, x, y^{'} )\} =[r(\lambda-\mu),U(\la )\otimes I+I \otimes  U(\mu )
] \delta(y-y^{'}), \nonumber \\
  r(\lambda-\mu)=  {P} { r_0(\lambda- \mu)}, \ P=\frac 1 2
(I+\sum_{a=1}^3 \sigma^a\otimes \sigma^a),  \ r_0= \frac 1 {2 (\la
-\mu)},
\ll{cybe}\eea
which is known as the classical Yang-Baxter equation (CYBE) 
with the rational $ r(\lambda-\mu)$-matrix along the relevant direction $y$ (trigonometric and elliptic
$r$-matrices are not relevant in the present context).  
For proving the integrability of the system at a global level together with
the sufficient condition (\ref{cybe}) one needs also the ultralocality condition
\be
\{U(\la, y  ), \otimes U(\mu, y^{'} )\} = 0, \mbox {at}
y \neq y^{'}
\ll{Ulocal}\ee
at different points on the $y$ axis, which follows also from  (\ref{cybe}).

Note, that  CYBE with the same $r$-matrix as in (\ref{cybe}), though along the
$x$ direction,  is valid   also 
for the  known LLE model \c{LLE}, which however  gives  much simpler relations 
 (involving  only 2 nontrivial relations)
compared to the present case, having 10 nontrivial relations, with few major
ones as
\bea 
\{U_{11}(\la,y),  U_{12}(\mu,y' )\} = 2(U_{11}(\mu ) - U_{11}(\la )) r_0 (\la-\mu)
\delta(y-y')
, \nonumber \\
\{U_{12}(\la,y),  U_{21}(\mu,y' )\} = (U_{12}(\mu ) - U_{12}(\la )) r_0 (\la-\mu)
\delta(y-y')
\ll{cybeij}\eea
etc.
 This happens   due
to much complicated structure of the present Lax operator (\ref{V}). However,
interestingly, all these involved  CYBE relations are satisfied
simultaniously due to the novel PB relations among the field components of
the present model  as in
(\ref{PB}), or in more elaborate form as

\bea 
\{S^3(y), S^3_x(y')\} = \frac i 2 
\delta(y-y'),  
\ \
\{S^+(y), S^-_x(y')\} =  i 
\delta(y-y')
, \ \{S^-(y), S^+_x(y')\} =  i 
\delta(y-y')
 \nonumber \\
\{S^3(y), S^-_x(y')\} = \{S^-(y), S^-_x(y')\} =\{S^-(y), S^+(y')\} = 0 
\ll{PBij}\eea
etc.
 
It is remarkable, that in spite of the presence of a $x$-derivative term
in the Lax matrix (\ref{V}),
 it satisfies the necessary ultralocality condition
(\ref{Ulocal}) due to the PBs (\ref{PBij}).  This is because not $x$ but  $y$ is the relevant direction
here, where the fields commute at space-separated points along  $  y$,
    reflecting the
quasi-2d nature  of our model with  space-asymmetry. Recall that the fields
in the standard LLE the related PBs hold for space points along the
$x$-direction. 

 Now we switch over to the quantum generalization of our  new integrable field model
and show that as a  quantum field model it passes the criteria of quantum
integrability and allows exact Bethe ansatz solution with intriguing
properties.     

\section{Quantum field model and exact solution}
For quantum generalization the recommended procedure is to lattice
regularize the fields by discretizing the space along the relevant direction
$ y \to j$  to obtain 
 ${\bf S}(x,y) \to {\bf S}_j(x), $ and  express the associated Lax operator of the model
(\ref{V}) in a  
discretized form:
 $U^j(\la )=I + \De \ U(\la, y \to j ),$ with explicit expression for its
matrix operator elements as
 
\bea
U^j_{11}(\la )= I+i \De u^j, \ U^j_{22}(\la )= I- i \De u^j, \ u^j= 
\frac {2} {\lambda^2} { S_j^3} +\frac {i} {2 \lambda}
 (S_j^-S^+_{jx}-S ^+_{j}S^-_{jx}) \nonumber \\
U^j_{12}(\la )= i \De(\frac {2} {\lambda^2} { S_{jx}^-} +\frac {1} { \lambda}
 (S_{j}^-S^3_{jx}-S_{j}^3S^-_{jx})), 
\ \ U^j_{21}(\la )= i \De(\frac {2} {\lambda^2} { S_{j}^+} +\frac {1} { \lambda}
 (S_{j}^+S^3_{jx}-S_{j}^3S^+_{jx})), 
\ll{Uj}\eea
where  $S^a_j, a=1,2,3 $ are now quantum field operators. 
Note that the lattice regularization is enough to perform here along the
$y$-direction  keeping the space variable $x$ to be continous, since the
Lax operator here is defined as a shift operator along $y$. Nevertheless, it
is to be noted, that the $(2+1)$ dimensional field ${\bf S}(x,y,t) $ 
 depends on the coordinates $x,y,t$, where the field ${\bf S} $ together
with
its $x$-derivatives enter in the Lax operator in a nonlinear form (see
(\ref{Uj})), with the   
lattice regularization needed  for  $y \to j $ only. This fact also exhibits a
quasi-2d dependence of our field with marked space-asymmetry. In fact the
space directions are scaled differently, which is acceprable for
nonrelativistic  models, as for example in the well known  $ (2+1)$-dimensional 
  integrable KP equation \c{KP}.
the Poisson brackets (\ref{PBij}) can be 
 quantized to  yield the commutation relations
 between the components of the  field as 
\bea
 [ S_{j}^3,  S_{kx}^3]=  \frac {\alpha } { \Delta} \delta_{jk}, \ 
[ S_{j}^-,  S_{kx}^+]= - \frac {2 \alpha } { \Delta} \delta_{jk}, \ 
[ S_{j}^+,  S_{kx}^-]=  \frac {2 \alpha } { \Delta} \delta_{jk}, 
 [ S_{j}^3,  S_{k}^{\pm}]= 
[ S_{j}^-,  S_{k}^+]= 0
\ll{CRs}\eea 
etc.

For showing the quantum integrability of the model, the operator elements of
the discretized quantum Lax matrix (\ref{Uj}) should satisfy certain
algebraic commutation relations, which can be given in a compact matrix form
by the quantum Yang-Baxter equation (QYBE)

\be    R(\la-\mu) \ U^j(\la ) \otimes U^j(\mu ) =
U^j(\mu ) \otimes U^j(\la ) R(\la-\mu),\ll{qybe}\ee
at each lattice site $j =1,2, \ldots N, $ 
  together with an {\it ultralocality} condition 
 \be  [U^j(\la )\otimes  U^k(\mu)]=0, \  \ j\neq k , \ll{qUlocal}\ee
Note, that these relations are quantum generalization of the classical
equations (\ref{cybe}, \ref{Ulocal}), where 
the quantum  $4 \times 4 \   R$-matrix with  nontrivial elements :
 \bea R^{11}_{11}=R^{22}_{22 }\equiv  a (\la-\mu) =\la-\mu
+i\alpha, \ R^{12}_{21}=R^{21}_{12} & &     \nonumber \\  \equiv  b (\la-\mu) =\la-\mu  ,
    R^{11}_{22}=R^{22}_{11}  \equiv   c=i\alpha, 
 & &  \ll{Rrat} \eea 
is a quantum extension of the the classical $r$ matrix appearing  in (\ref{cybe}). 
It is to be noted, that in spite of the presence of
a $x$-derivative  term in the quantum Lax operator (\ref {Uj}), thanks to
the   new CRs (\ref{CRs})  the necessary
ultralocality condition (\ref{qUlocal}) holds. This is because    
$y$  and not $x$ is  the  concerned direction  here,, where the fields  commute
    at space separated
points  along   $y \to  j $. 

If we define a global operator  for $N$-lattice sites as 
$T(\la)=\prod_{j=1}^N U^j(\la ) , $
 through the  lattice regularized quantum Lax operator $U^j(\la ), $ which
satisfies the QYBE (\ref{qybe}) together with (\ref{qUlocal}), then the
global  monodromy operator $T(\la)$ 
must also satisfy the QYBE
 \cite{fadrev}
\be    R(\la-\mu) \ T(\la ) \otimes T(\mu ) =
T(\mu ) \otimes T(\la ) R(\la-\mu),  
\ T(\la ) = \left( \begin{array}{c}
 A(\la ), \  \ \quad 
  B(\la )  \\ 
  B^\dagger(\la ) ,   \quad \ \ 
A^\dagger(\la )
          \end{array}   \right),
\ll{qybeg}\ee
 with the
same $R(\la-\mu)$-matrix.  This happens due to the coproduct property of the underlying
Hopf algebra , which keeps an algebra invariant under its tensor
product \cite{chari}.  This global QYBE (\ref{qybeg}) serves two important purposes.  First, it proves
the quantum integrability of the model by showing the mutual commutativity
of all conserved operators.  Second, it derives the commutation relations
between the operator elements of $T(\la )$, which are used for the exact
algebraic Bethe ansatz  solution of the EVP.

In more details: multiplying  QYBE 
(\ref{qybeg}) from left by $R^{-1} $,  taking the trace from both  sides  and
 using the property of cyclic rotation of matrices under
the trace,  
one can show   that $ \tau (\la) = {\rm trace} \ T(\la) $  commutes:
 $[\tau (\la),\tau
(\mu) ]=0. $  This in turn   leads to the Liuoville integrability condition:
 $[C_n,C_m]=0, \ n,m=1,2, \cdots $, 
  since the   conserved set of  operators are generated from $\ln
 \tau(\la)=\sum_j C_n \lambda ^{n}, $ through expansion in the spectral
parameter $\la  $.  Following this construction and
 exploiting the explicit form of the Lax matrix (\ref{Uj}),
  we can derive, in principle, 
 all conserved operators $C_n, \ n=1,2,\ldots $ for our model, as given for
the classical case in (\ref{Cm21012}).

Therefore, for proving the quantum integrability of the  proposed field model,
associated with the quantum Lax operator (\ref{Uj}), we have to satisfy the
QYBE (\ref{qybe}) for each matrix elements.  However due to the quadratic
spectral power dependence  of the Lax operator together with its
 nonlinear dependence on the
fields and 
 its more complicated  structure, 
the problem becomes much harder compared to the known quantum LLE model
\c{qLLE}. 
However  all these relations (as we see below in explicit form) are
satsisfied due to the new quantum commutation relations (\ref{CRs}) for our quantum
field, upto  order $O(\Delta) $, which however
 is enough for quantum field models obtained at
$\Delta \to 0 $.

Note that comparing with the well known quantum LLE model, 
where  only $two$ nontrivial relations
appear in the QYBE, the present model  brings harder challenges, since in
total $ten$ nontrivial  quantum equations arise  in its QYBE, as we discuss below.
 
\subsection{QYBE for the integrable field model}

In  QYBE (\ref {qybe}) with $R$-matrix (\ref{Rrat}), for our quantum
integrable  model  we insert the associated    discretized quantum Lax matrix  $
U^j$ as  in   (\ref{Uj}) and look explicitly for the validity of QYBE relations
for  each of the matrix operator element. We find, that  out of  total 16  operator
relations,  except 4 diagonal and 2 extreme
off-diagonal terms,  all the   other 10    relations $Q^{ij}_{kl} $  stand nontrivial and
 their validity needs to be proved using in particular  the  operator product
relations at the coinciding points:
\bea
  [S_{j}^3,  S_{jx}^3]=  \frac {\alpha } { \Delta} , \ 
[ S_{j}^-,  S_{jx}^+]= - \frac {2 \alpha } { \Delta} , \ 
 [ S_{j}^+,  S_{jx}^-]=  \frac {2 \alpha } { \Delta}, 
 \ \ \mbox{and} \ \  [ S_{j}^3,  S_{k}^{\pm}]= 
[ S_{j}^-,  S_{k}^+]= 0,
\ll{CRop}\eea 
at  space-seperated points, following from the CR 
(\ref{CRs}).

Using the expressions for $a(\la -\mu), b(\la -\mu), c $ from (\ref{Rrat}) and   CR
(\ref{CRop}) we may check the validity of  \bea 
Q^{11}_{12} 
 =a \ {U^j}_{11}(\la ) {U^j}_{12}(\mu )- b \ {U^j}_{12}(\mu ){U^j}_{11}(\la
)   - c \  
{U^j}_{11}(\mu ){U^j}_{12}(\la ))  = 
\ \   +O(\De ^2) =0,
\ll{App_Deqybe} \eea 
 upto order $O(\De ^2)$.
Similarly, one proves the conjugate relations $Q^{11}_{21}, Q^{21}_{11}, Q^{12}_{11}$
and similar relations $Q^{22}_{12}, Q^{22}_{21},$ $ Q^{12}_{22}, Q^{21}_{22} .$

The  remaining two relations can also be  proved
 with the use of the same operator product relations   (\ref{CRop}): 
\bea Q^{12}_{21} 
 =b \ [{U^j}_{12}(\la ), {U^j}_{21}(\mu )]  \ \   +c \  ({U^j}_{22}(\la ){U^j}_{11}(\mu
)\ \  -
{U^j}_{11} (\la ){U^j}_{22} (\mu ))  \   = 
\   =0 , 
\ll{App_exactqybe}\eea
which holds
 exactly in all orders of $\De $ and
similarly for the conjugate relation $Q^{21}_{12}$.
This proves thus the
validity of all QYBE relations for our quantum quasi-2d NLS field model, associated
with the higher Lax operator (\ref{Uj}) and algebraic relations
(\ref{CRop}), obtained at the limit   $\De \to 0 $ .

\section{ Algebraic Bethe ansatz  for the eigenvalue problem}
As noted above,   the
 monodromy operator  $T(\la)$ associated with  our quantum Lax operator
(\ref{Uj}), 
 as guaranteed by QYBE (\ref{qybe}) together with the ultralocality
condition  (\ref{qUlocal}),
 satisfies also the same  QYBE (\ref{qybeg}) with the rational $R$-matrix (this is due to
the  Hopf algebra property \c{chari} inherant to this peoblem ). Therefore,
 we can follow the procedure
for the algebraic BA, close to the formulation  of  the 1d quantum
LLE model \c{qLLE}.  As we have discussed above, $ \tau( \la )={\rm
trace}  T(\la)=
 A(\la ) + A^\dagger (\la )$ is linked to the generator of the conserved
operators $C_n, \ n=1,2, \ldots $, including the Hamiltonian (\ref{H}). 
The off-diagonal elements of
 $T^{12}(\la )=B(\la)$ and $T^{21}(\la )=B^\dagger(\la)$,  on the other hand,
 can be
considered as   generalized {\it   creation} and {\it   annihilation}
operators, respectively.
For solving  the eigenvalue problem (EVP) for all conserved operators:
$C_n|M>=c^M_ {n}|M>, \ n=1,2, \ldots $  simultaneously,
we  construct exact M-particle Bethe state  $|M>= 
B(\mu_1)B(\mu_2)\cdots B(\mu_M)|0>, $  on a  pseudo-vacuum $|0> $ with the
property \  \ 
 $B^\dagger (\mu_a)|0> =0, \  A(\la)|0>= g(\la ) |0>, $ \ where 
numerical function $ g(\la )$  depends on the vaccum expectation value of the Lax
operator: $U_0(\la)=<0|U^j(\la)|0> $    and aim to solve 
the EVP:\ \  $\tau (\la )|M>= \Lambda_M (\la, \mu_1, \mu_2,\ldots,  \mu_M )
|M>, \ \  $
with exact eigenvalues   $\ln \Lambda_M (\la, \{ \mu_a \})=\sum_j c^M_n ( \{ \mu_a \})
 \lambda ^{n}$.

\subsection{Exact solution for  quasi 2d quantum   field  model}
For obtaining the final  result for our quantum  field model, on infinite
space interval, 
we have to switch over to the field limit: $\De \to 0 $ with total lattice site
  $N \to \infty $ and then take the interval $L=N \De \to \infty, $ 
assuming vanishing of the  field $S^\pm_{j }
 \to 0,  \ S^3_j \to 1, \  \mbox {at  }\  j \to \infty $, 
 compatible with  the natural condition of having the
vacuum state  at  space   infinities, yielding the asymptotic Lax matrix
 $U^{j}(\la )|_{j\to \infty}=U_0(\la )= I+\frac {2i } {\lambda^2} \Delta \sigma
^3.$ 
Therefore, we have to  shift over  to the  monodromy  matrix at the field
limit  defined as
\be T_f(\la )=U_0^{-N} \ T(\la )  \  U_0^{-N}, \ \ \  N \to \infty , \ll{Tf} \ee
and for  further construction  introduce 
 $\ \ V(\la,\mu) \equiv U_0(\la )\otimes U_0(\mu ) , \ $
 $W(\la,\mu)=(U^j(\la )\otimes
U^j(\mu ))_{j \to \infty} .$
  We may   check from the QYBE (\ref{qybe}) that $W $ satisfies the relation
   $ \ R(\la - \mu) W(\la, \mu)= W(\mu, \la) R(\la - \mu),  \ $
using which  we can derive  from  QYBE (\ref{qybeg}),
 that the  field monodromy   matrix (\ref{Tf}) also satisfies the QYBE
\be    R_0(\la,\mu) \ T_f(\la ) \otimes T_f(\mu ) =
T_f(\mu ) \otimes T_f(\la ) R_0(\la,\mu),
\ll{qybegf}\ee
but  with a  transformed $R $-matrix: 
\be R_0=S(\mu,\la) R(\la -\mu) S(\la, \mu), \ 
S(\la, \mu)=W^{-N}V^N,   \  N \to \infty , \ll{R0} \ee  where
 $R(\la -\mu)$ is the original rational quantum  $R -$matrix (\ref {Rrat}) 
(see \c{fadrev}  for similar details on  1d NLS model).
Based on  the above  formulation, using the field operator products:
$S^+_jS^-_{j,x}= 2\frac{ \alpha} \De, \ S^-_{j,x} S^+_j=0,   $ at $j \to
\infty , $   
we can calculate  explicitly  the relevant objects needed for our field model. In
particular,
the  central $2 \times 2  $  block $W_c $ for 
matrix $ W $ turns out to be 
\bea W_c(\la, \mu)= I+\De \ { M}(\la, \mu)
\left(\begin{array}{cc}   (\la - \mu)   &   
0 \\ 
-2 \alpha & - 
 (\la - \mu)
\end{array}\right), 
 \ll{W} \eea
with an intriguing factorization of its spectral dependence  by a
prefactor $ { M}(\la, \mu)=2 \frac {(\la + \mu)} {\la ^2 \mu ^2} , $
 which is  the key reason behind the success
of  the  exact algebraic Bethe ansatz solution for our  field model, inspite of
the  more complicated form of  its Lax operator with nonlinear dependence on
the spectral parameter and on the fields.

For constructing $R_0$  using  definition (\ref{R0}), we have to find first
the matrix $S(\la,\mu) $, taking proper limit of $W^{-N} $ at $ L \to \infty
$ using (\ref{W}) .  Through some algebraic manipulations, which are skipped
here, we finally arrive at the field limit,
 to a  simple  form for  $R_0$ matrix, expressed through its
nontrivial elements as
 \bea R^{11}_{11 }=R^{22}_{22 }=  a (\la-\mu) 
,   R^{12}_{21}=    b (\la-\mu) , \     R^{11}_{22}=R^{22}_{11}  =0,
   & & \nonumber \\ 
  R^{21}_{12} = b (\la-\mu)  - \frac
{\alpha ^2} {\la-\mu } + \frac
{\alpha ^2 \pi} {M(\la , \mu) } \delta(\la - \mu), & &     
   \ll{R0jk} \eea 
where $ { M}(\la, \mu)=2 \frac {(\la + \mu)} {\la ^2 \mu ^2} ,  \ $
 $ a(\la-\mu), b(\la-\mu)$  as in  (\ref{Rrat}) and the $\delta (\la -\mu) $ term
vanishes at $\la \neq \mu  $.  
It is intereesting to compare (\ref{R0jk}) with the original quantum $R$-matrix
(\ref{Rrat}). 
Now from  QYBE (\ref{qybegf}) relevant for the field models, we can derive  using   the
$R_0$ matrix (\ref{R0jk}), the
required CR between the operator elements
 of $T_f (\la) $. 
In particular, we get for our quantum field model  the commutation relation 

 \be A_f(\la) B_f(\mu_a)=(f_a(\la -\mu_a)-  \frac
{\alpha ^2 \pi \la ^2 \mu ^2} {2(\la + \mu_a) } \delta(\la -\mu_a))
 B_f(\mu_a ) A_f(\la),  \ll{AB} \ee 
 where $ \ f_a= \frac 
{\la -\mu_a - i \alpha } {\la -\mu_a
 }. $  Note that at $\la  \neq \mu_a $, the singular term
with a  prefactor bearing the imprint of
the  $\la ^2 $ dependence of our Lax matrix, vanishes and  the relations 
   coincide in parts  with those of  the  known LLE  model, though
only formally, since the nature of the basic fields is completely different
for these two models. 

Using this result and the property of the vacuum state: $A_f|0>=|0> , $ 
we obtain  the   exact  EVP for 
\be A_f(\la)|M> =  F_M |M> , \ \ \mbox{ as}  \ F_M = \prod_a^M f_a(\la
-\mu_a), \ \ A_f(\la)|0>=|0>\ee and hence for $\tau_f(\la ) $, 
 which yields  finally the exact  eigenvalues
$c^{(M)}_n $  for 
conserved operators $C^{(M)}_n $ from the relation
\be  \tau_f(\la)|M> =  \Lambda_M (\la )|M>, \ \ \ln \Lambda_M (\la )= \sum_n\
c^{(M)}_n \la^n, 
\ll{tauCn}\ee
all of which can be extracted systematically. 
 Few lower ones from this infinite series
   take the explicit form

\bea 
 c^{(M)}_0&=& \sum_{a=1}^M \rho_0(\mu_a), \ \rho_0(\mu_a) =
   \frac 1 2 \ln (1+ \frac {\alpha^2} {\mu_a^2}),, \ \ 
   c^{(M)}_1= \sum_{a=1}^M \ \rho_1(\mu_a) , \ \rho_1(\mu_a) = \frac { 2 \alpha^2} {\mu_a(\al ^2 +
\mu^2_a)}) \nonumber \\ c^{(M)}_2&=&  \sum_{a=1}^M \  \rho_2(\mu_a),
\ \rho_2(\mu_a)
 = \alpha^2 \left [ \frac {3\mu^2_a+ \al ^2 } 
{(\mu_a(\al ^2 +
\mu^2_a))^2} \right ] , \nonumber \\ c^{(M)}_3&=&\sum_{a=1}^M \
\rho_3(\mu_a), \ 
\rho_3(\mu_a)=  2 \alpha^2
  \left [ \frac { \mu^4_a + 3 \al ^2 \ \mu^2_a+ \al ^4 } 
{ 3 (\mu_a(\al ^2 +
\mu^2_a))^3} \right ]  \ll{C012} \eea 
etc. where $ H=C_2  $  is the  Hamiltonian of  our model. Therefore  we obtain 
 the exact energy spectrum as  $E_M=c^{(M)}_2 $, for the M-particle
scattering state, which clearly differs from that of the known LLE model
\c{fadrev,qLLE}. However, the overal spectrum of the conserved operators
coincides in both these models due to the same quantum $R$-matrix involved
in both these cases. Note, that due to the vaccum state property  $A_f|0>=|0> , $
the imprint of the Lax opeartors, which are widely  different for the LLE
and the  present model, 
 is lost at the field limit, leaving the $R$-matrix as the  determining
factor for the eigenvalues of the conserved operators.

It is interesting to compare the eigenvalues of the 
conserved operators (\re {C012}) and their corresponding classical expressions
(\re{Cm21012},\re{H}) .
  It is remarkable, that in spite of the highly
nonlinear field interactions present in the Hamiltonian (\re{H}), the scattering
spectrum shows no coupling between individual quasi-particles, mimicking a
free-particle like scenario.

On the other hand,  the bound-state or the quantum {\it soliton} state, which
is  obtained
for the complex   {\it string} solution for the particle  momentum:
  $ \mu^{(s)} _ a =\mu _0 +  i \frac \alpha 2 ( (M+1)-2a), \ a=1,2, \cdots,
M $
where $\mu _0 $ is the average particle  momentum 
 and  $ \alpha $ is the coupling constant, induces  mutual interaction
   between the particles.
 Recall, that  a bound-state
becomes stable, when its energy is lower than the sum of the
individual free-particle energies with the average momentum, which in turn is ensured by the {
negative}
values of the binding energy.  More negative binding energy
indicates more stable bound-states.
  The corresponding bound state energy spectrum can   be calculated 
for the present model for the  $M > 1$-particle bound-state, though it
becomes rather cumbersome due
to complicated  expression of $c^{(M)}_2 $ involving series sum of 
rational functions due to the rational dependence of the energy density $  \rho_2(\mu_a) $  on
$\mu_a $. Though it is straightforward  to extract the bound state energy,
the resulting  expression is lengthy,   
containing several  terms involving  polygamma functions and will not be
reproduced here for the general case of $M>2 $ . However  
To illustrate the situation and to demonstrate the intriguing stability
condition for the bound state of the model we present only the simplest case
for the  energy of the
$2$-particle bound state given by the following expressions containing both 
positive $E^+ $ and negative $E^- $  contributions:  
\bea E_2^{(s)}&=& E^+-E^-   
\nonumber \\
E^+&=& \frac {1} {4} ( 16  \alpha^3 \ + \ 8 \alpha^2\ \mu_0^2 + \frac {48
\alpha^2}{(
\alpha^2 + 4 \mu_0^2)^2} + \frac{(
   432 \alpha^2  + 2176 \alpha^7 )} {(9 
\alpha^2 + 4 \mu_0^2)^2}  \nonumber \\ & & + 
 \frac{(
   12 + \alpha^4 (32 + 1041 \alpha^2 ))}{(9 \alpha ^2 + 4
\mu_0^2)})
\ll{E+}\\
E^-&=& \frac {1} {4}  (66 \alpha ^4   +\frac {12}{(
   \alpha ^2 + 4 \mu_0^2)} + \frac{( 320 \alpha ^6  + 4025 \alpha ^8)} 
{(9 \alpha ^2 + 4 \mu_0^2)^2} + \frac {384\alpha ^5} { (9\alpha  ^2 + 4 \mu_0^2)})
 \ll{E-} \eea
Note that when the $2$-particle bound state energy $ E_2^{(s)} $ for our
 quai-2d quantum field model becomes less than the sum of the energies:
$2 \ \rho_2(\mu_0) $ of
two 
free particle scattering state, the bound state becomes stable due to the
nontrivial  value of the of the binding energy, which would be determined by
the competing contributions of the positive  (\ref{E+}) and negative
(\ref{E-}) parts of the bound state ebergy.          


\section {Concluding remarks and Outlook}

Summarizing the saliant points of our construction we  note, that since both
the standard $(1+1)$-dimensional LLE model and the present quasi
$(2+1)$-dimensional model in the quantum case are linked with the same
$R$-matrix and the eigenvalues of the conserved operators are determined
mainly by its $c$-number matrix elements, especially at the field limit, the
eigenvalues coincide formally for the higher conserved operators in both the
above models, although the energy spectrum corresponding to different
Hamiltonians for these models are distinct.  At the same time, though these
two models are intimately related, the contents and the structure of these
models are widely different, with different nature of their basic fields. 
The fields ${\bf S}$ in the known LLE model behave like spin fields
satisfying the $su(2)$ algebraic relations (\ref{CRlle}) and exhibits an
important constraint ${\bf S}^2=I$, as a Casimir operator.
 The associated  Lax operator $
U_{lle}$ (\ref{Ulle}), 
generating shift along the $x$-direction, has only linear dependence on the spectral parameter
 and on the fields and satisfies as an ultralocal model the quantum Yang-Baxter equation (QYBE)
along the $x$-axis.  On the other hand, the basic fields of the present
model satisfy  commutation relations (\ref{CRs}), which do not allow any
contraint and  the fields behave like three independent real scalar
fields with nonstandard commutators, exhibiting unusual and significantly
different nature of the fields.  These novel CRs, involving
 $x$-derivative of the field  are defined 
along the $y$ axis, showing
the  quasi
$(2+1)$d  character of the model, which  is reflected  also     
in the form of its conserved quantities (\ref{Cm21012}).  The related  Lax operator $
U(\la)$ (\ref{V}),  representing  infinitesimal shift operator
along the transverse direction $y$,  has a  nonlinear dependence on the spectral parameter
as well as on   the  fields and  contains  $x$ derivative of the
field, showing higher scaling order  and 
 space-asymmetry of the model. In spite of these explicit unfavorable facts the quantum Lax
operator of our model  satisfies the crucial   ultralocality condition  and the 
QYBE 
with the rational quantum $R$-matrix,  
with $y$ as the relevant direction, thanks  to the unusual CRs of the
fields.

The  integrable model, proposed here, is important from several point of view.   
First, as a {\it new} integrable quantum field  model satisfying the 
QYBE  and exactly solvable by the algebraic Bethe
ansatz, is important by its own
right. Second, aa a  quantum  field model built     in quasi 2-dimensions, 
 going beyond the standard construction  of the  existing 
1-dimensional  quantum integrable models   and solved exactly by the Bethe ansatz,
is a significant achievement.
Third, as a  quantum integrable model, constructed
following the idea of higher order Lax operator,  provides
a  nontrivial example of another new model in quasi 2-dimensions,
 needed for  proving the conjecture and showing the universality 
of the approach  proposed in  \c{KundNPB15}. Fourth, since the quantum LLE received
renewed attention  due to its link with the string theory, following the ADS/CFT
correspondence, the quantum field model proposed here, due to its close proximity
with the quantum LLE, could also  be interesting from  other angles.

The idea folloed here should show the path in constructing a novel class of
higher-dimensional field and lattice models both at the classical and the
quantum level and should help in discovering new type of algebraic
relations, like those found here.

\end{document}